\documentclass{article}
\usepackage{dcase2018,amsmath,graphicx,url,times,booktabs, tabularx,mathtools, float, fixltx2e}

\title{sound event detection using Weakly labelled semi-supervised data with
GCRNNs, VAT and self-adaptive label refinement}

\twoauthors
  {Robert Harb}
    {Graz University of Technology, Austria\\
     robert.harb@student.tugraz.at}
  {Franz Pernkopf}
    {Graz University of Technology, Austria\\
     Signal Processing and Speech Communication Laboratory \\
     pernkopf@tugraz.at}

\begin{document}

\ninept
\maketitle

\begin{sloppy}

\begin{abstract}
In this paper, we present a gated convolutional recurrent neural network based approach to solve
task 4, large-scale weakly labelled semi-supervised sound event detection in domestic environments,
of the DCASE 2018 challenge.
Gated linear units and a temporal attention layer are used to predict the onset
and offset of sound events in 10s long audio clips.
Whereby for training only weakly-labelled data is used.
Virtual adversarial training is used for regularization,
utilizing both labelled and unlabelled data.
Furthermore, we introduce self-adaptive label refinement, a method which allows unsupervised adaption of
our trained system to refine the accuracy of frame-level class predictions.
The proposed system reaches an overall macro averaged event-based F-score of $34.6 \%$,
resulting in a relative improvement of $20.5 \%$ over the baseline system.
\end{abstract}


\section{Introduction}

In this paper we summarize the methods we use to solve task 4 \cite{2018arXiv180710501S} of the DCASE 2018 challenge,
the \textit{large-scale weakly labelled semi-supervised sound event detection in domestic environments}.
In contrast to audio tagging (AT), sound event detection (SED) not only requires to detect the
 presence of an event, but also a prediction about the temporal location in a given audio recording.
Whereby in the data provided by the DCASE challenge, one input sequence  possibly contains multiple
occurrences of different event classes with potential temporal overlaps.
Additionally, the training data is only weakly labelled. Therefore for
training, the labels of each clip contain only information about the presence or absence of an event,
but no strong labels which indicate the exact temporal onset and offset.

The proposed method uses a gated convolutional recurrent neural network (GCRNN).
This is similar to the best model of last years DCASE 2017 challenge task 4 \cite{xu2017large} which also used a GCRNN based approach.
Although, the objective of the 2017 and 2018 DCASE challenge is SED, there are significant differences
in the structure of the provided training data and evaluation metric.
More precisely, the following changes have been made at the 2018 challenge:
\begin{itemize}
\item The amount of weakly labelled training data is significantly smaller, 1,578 compared to 51,172.
\item In addition to the weakly labelled training set, there are unlabelled in-domain and unlabelled out-of-domain sets provided.
\item The domain of the events is different: \textit{domestic environments} compared to \textit{smart cars}.
Whereby the number of classes decreased from 17 to 10.
\item For evaluation, an event-based F-score with a 200ms collar on onsets and offsets is used,
instead of a segment-based error rate which is determined of one-second segments.
\end{itemize}

With our work we show that a GCRNN based approach for SED similar to \cite{xu2017large},
is also suitable in a setting with the aforementioned differences.
Whereby we introduce two major changes:

First, to incorporate the provided unlabelled data we use virtual adversarial training (VAT) \cite{vat}.
VAT has, amongst others, already been used successfully in semi-supervised text \cite{vattext},
image classification \cite{vat}, acoustic event detection \cite{pernk} and phone classification \cite{3420} tasks.
Furthermore, VAT showed competitive performance
against other deep semi-supervised learning algorithms \cite{oliver2018realistic}.

Secondly, as an extension to the attention mechanism we introduce an algorithm we call self-adaptive label refinement (SALR),
which uses unlabelled input data and clip-level class predictions to refine the frame-level predictions of our model.

\section{Proposed Method}

\begin{figure}[h!]
  \centering
  \centerline{\includegraphics[width=\columnwidth]{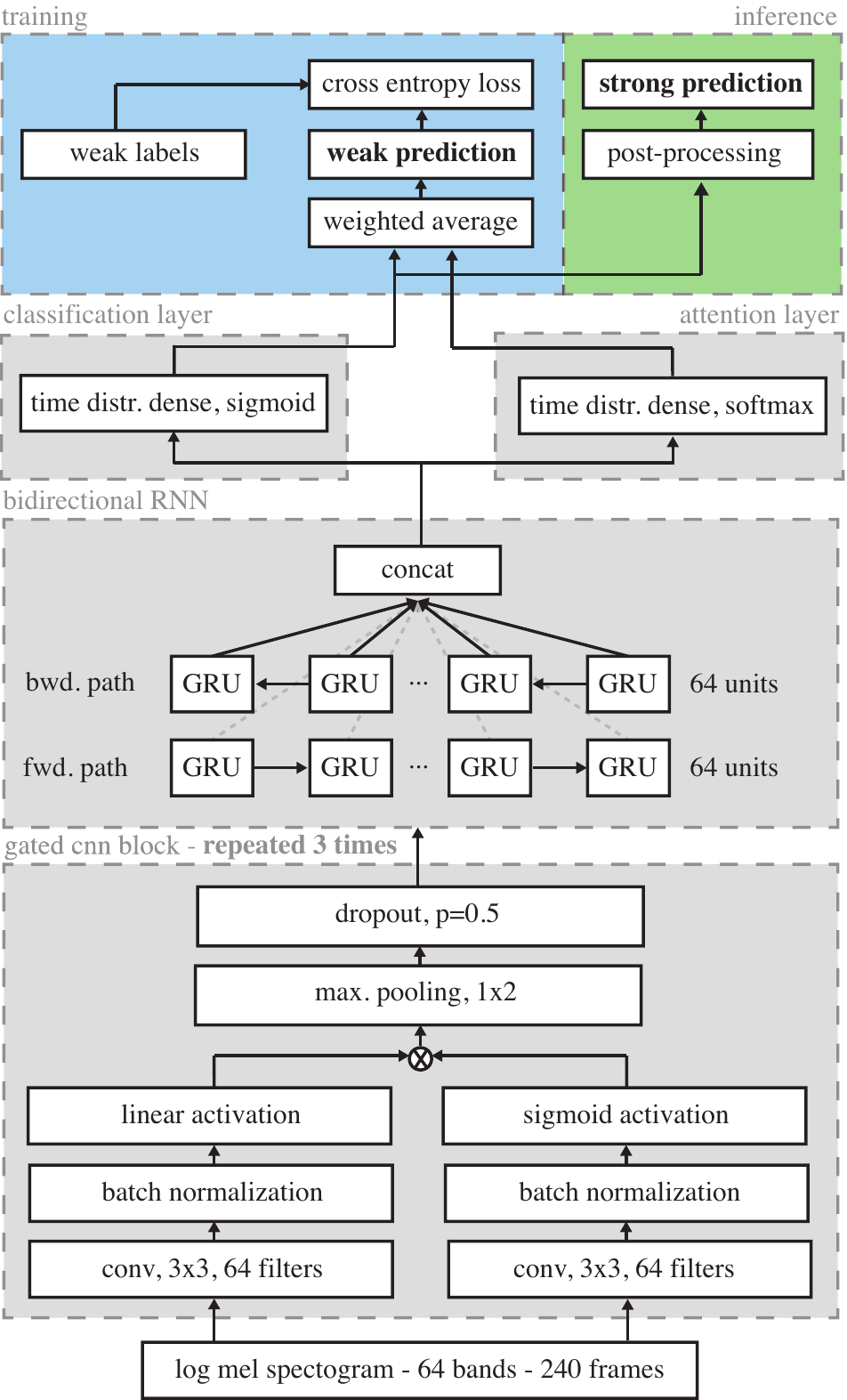}}
  \caption{Network structure}
  \label{fig:network}
\end{figure}

\subsection{Gated convolutional recurrent neural network}
The winning team of last year's DCASE SED task \cite{xu2017large} showed that
using gated linear units (GLUs) \cite{gated_units} instead of commonly
used activation functions like rectified linear units (RELUs) or leaky ReLUs
in the CRNN is a useful approach for SED.

Gating mechanisms have been used successfully in a variety of neural network architectures.
For example in RNNs using LSTM \cite{lstm} cells,
which have a separate input, output and forget gate.
The rough idea behind gating mechanisms is to have a gate which can control how information
flows in the network.

In the setting of SED, the GLU units
 should adapt their behaviour such that they act as an attention mechanism
 on the time-frequency (T-F) bin of each feature map.
 They can set their value close to one if information related
 to any of the considered audio events passes through,
 and otherwise block the flow of unrelated information by setting their value close
 to zero.

\noindent GLUs are defined as follows:
\begin{equation}
\mathbf { Y } = ( \mathbf { W } * \mathbf { X } + \mathbf { b } ) \odot \sigma ( \mathbf { V } * \mathbf { X } + \mathbf { c } ),
\end{equation}

\noindent where $\mathbf{W}$ and $\mathbf{V}$ denote the convolutional filters with their respective biases $\mathbf{b}$ and $\mathbf{c}$,
$\sigma$ is the sigmoid function, $\mathbf{X}$ denotes the input to the layer, and $\odot$ denotes elementwise multiplication.

Figure \ref{fig:network} shows how the gated CNN blocks are incorporated into
the network, whereby in our model we use three subsequent gated CNN blocks.

The gated CNN blocks are followed by a bidirectional RNN containing 64 units in the
forward and backward path, their output is concatenated and passed to the
attention and classification layer which are described in Section 2.3.

The final prediction $y_c$ for the weak label of class c is determined
by the weighted average of the element-wise multiplication of
the attention and classification layer output of class c:

\begin{equation}
y_c = \frac{ \sum_t^T \mathbf{z}^{cla}_c(t) \odot \mathbf{z}^{att}_c(t)}{\sum_t^T \mathbf{z}^{att}_c(t)},
\end{equation}

\noindent where $\mathbf{z}^{cla}_c$(t) and $\mathbf{z}^{att}_c$(t) are the outputs of the classification layer and of the attention
 layer of class c. $T$ denotes the frame-level resolution of the input spectrogram, which
 is equal to the resolution of $\mathbf{z}^{cla}_c$(t) and $\mathbf{z}^{att}_c$(t), and t is the frame index.

\subsection{Virtual adversarial training}
We make use of VAT \cite{vat} for regularization.
We calculate the virtual adversarial loss such that the robustness of the model's posterior distribution
of predictions at clip-level $p(\mathbf{y}|\mathbf{x})$
is increased for small and bounded perturbations of the
log-scaled mel-spectrograms $\mathbf{x}$.

The adversarial perturbation $\mathbf{r}_{v\textnormal{-}adv}$ is computed by maximizing a non-negative
distance function between the unperturbed $p ( \mathbf { y } | \mathbf { x } ; \theta)$
and perturbed $p ( \mathbf { y } | \mathbf { x } + \mathbf{r} ;\theta)$ posterior.
Whereby $\theta$ denotes the current model parameter.
The Kullback-Leibler divergence KL is used as distance function between $p (  \mathbf{y}  |  \mathbf{x}  ; \theta)$ and $p (  \mathbf{y}  |  \mathbf{x}  + \mathbf{r} ; \theta )$,
and $||\mathbf{r}||$ is limited to the sphere around $\mathbf{x}$ with some radius $\leq \epsilon$,
i.e. $\mathbf{r}_{v\textnormal{-}adv}$ is determined as

\begin{equation}
\mathbf{r}_{v\textnormal{-}adv} = \arg \max _ {\mathbf{r}, \| \mathbf{r} \| \leq \epsilon }  \textnormal{KL} [ p (  \mathbf{y}  |  \mathbf{x}  ;\theta) | | p (  \mathbf{y}  |  \mathbf{x}  + \mathbf{r} ; \mathbf{\theta} ) ].
\end{equation}

There is no evident closed-form solution for $\mathbf{r}_{v\textnormal{-}adv}$, but \cite{vat} gives
a detailed derivation how to calculate an approximation of $\mathbf{r}_{v\textnormal{-}adv}$.
When using VAT the following additional cost is added to the objective function:

\begin{equation}
\mathrm { \textnormal{KL} } [ p ( \mathbf{y} |  \mathbf{x}  ;  \theta   ) | | p ( \mathbf{y} |  \mathbf{x}  +  \mathbf{r}  _ {  v\textnormal{-}adv  } ; \theta ) ].
\end{equation}

\noindent Since calculating the virtual adversarial perturbation only requires input $\mathbf{x}$ and does not require label $\mathbf{y}$,
VAT is applicable to semi-supervised training. Therefore we use it to incorporate
the unlabelled in-domain dataset into training.
However, we decided not to include any of the provided unlabelled out-of-domain data since it has been shown previously that
adding unlabelled data from different
classes than the labelled data, can actually decrease
the performance of semi-supervised learning algorithms like VAT \cite{oliver2018realistic}.

\subsection{Attention mechanism}
To predict the temporal locations of each audio event which is presented
in a given input sample, we use a similar approach as used in
\cite{xu2017large}. We extend it by introducing self-adaptive label refinement based on
weak and strong prediction alignment. This selects for each event class an
appropriate post-processing on the networks attention output.
In the following the term weak prediction is used to refer to predictions
at clip-level and strong prediction is used to refer to class predictions
at frame-level.

As depicted in Figure \ref{fig:network}, the output of a bidirectional
RNN is fed into both an attention and a classification layer.
The classification layer uses a sigmoid activation function
to predict the probability of each occurring class at each timestep.
While the attention layer uses a softmax activation over all classes.
Intuitively, using a softmax in the attention layer should aid the network to
learn to pick the most dominant class at each frame.
Although this might not be an ideal approach if temporal overlaps of multiple events are occurring,
since then a more dominant event might be able to suppress the activation of another one.

Figure \ref{fig:output} shows the output of the classification and attention layer
for one audio clip of the development set containing several events labelled as dog.
It can be seen that there is a clear correlation between ground truth
event labels and the activations of the attention and classification layer.
However it is not obvious how to extract the exact start and end points of
each individual event from the layer activations.
Our experiments showed that just taking the product
 of the attention and classification layer activations, thresholded with a fixed
 value for all classes, e.g. $0.5$, gives unsatisfactory results.
Also it has been shown in similar weakly labelled SED settings
that the trained network adapts differently for
different classes \cite{McFee2018AdaptivePO}. Especially there seems to be
 a difference between classes which tend to have short event durations in contrast to
classes which span the majority of timesteps of a clip.
Considering this, it might be necessary to use a different
post-processing
for each class on the networks attention output to account for that.
 The fact that no strong event annotations are available for training
 makes this a non-trivial problem, otherwise a simple approach
 would be to test several post-processing methods and select for
 each class the one which gives the best performance.

\begin{figure}[]
  \centering
  \centerline{\includegraphics[width=\columnwidth]{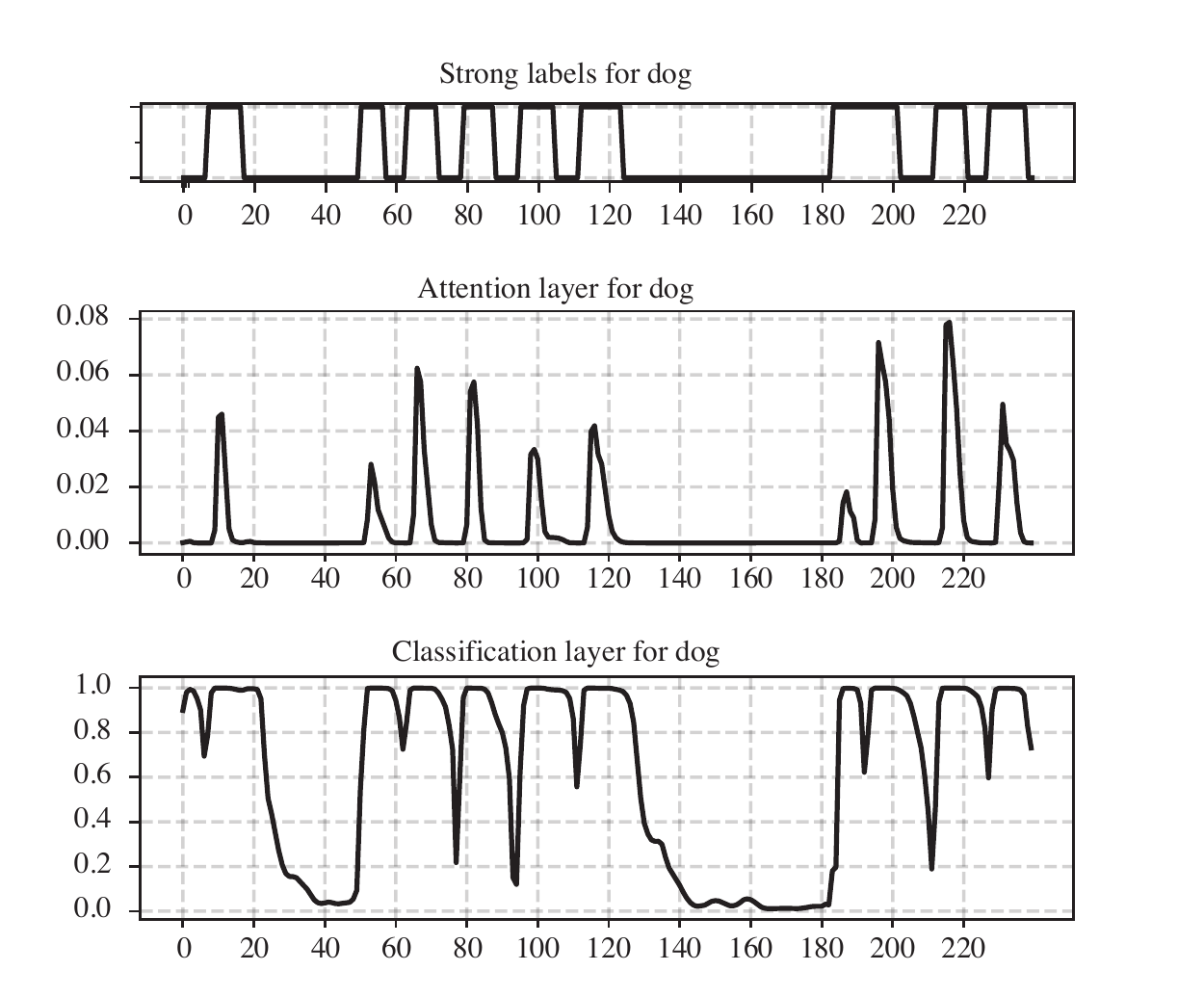}}
  \caption{Classification and attention layer activations for file: \textit{Y0a8RB5eOGJ4\_30.000\_40.000.wav} and class dog.}
  \label{fig:output}
\end{figure}

 \subsection{Self-adaptive label refinement (SALR)}

We introduce self-adaptive label refinement,
where we check the alignment between
strong and weak predictions, and use this as an
approximate prediction how well
a given
post-processing method performs at extracting the right onset and offset of events.
Using this approach we can use unlabelled data
 to estimate how well a given post-processing parameterization performs for each class,
 and take the best performing parameterization for our final strong prediction.

For post-processing we threshold the output value of the classification layer, followed by
a median filter. Therefore the parameters we vary in each iteration
are the threshold, and the width of the median filter.
However it should be noted that many other methods for post-processing are possible,
e.g. a second neural network which maps between the attention layer of the first network
and strong predictions might be a potential approach.

In particular, when training has finished, self-adaptive label refinement
repeats the following steps on each class with different post-processing parameterizations:

\begin{enumerate}
\item A full forward pass is performed to create weak and strong
predictions for each clip.
Whereby the following steps are only carried out for clips where
the weak prediction indicates occurrence of the current class.
\item
Using the strong predictions, the spectrogram of each clip
 is split up into two groups of new samples.

Each single event occurrence of the examined class
is extracted into new samples containing only
the temporal frames of the spectrogram which possibly contains
the event.
Those new samples are labelled with $1$.

 Additionally,
 another sample is created which contains only the temporal frames
 of the original spectrogram where no occurrence of the
 given class was predicted.
 Those are all labelled to 0.

\item The generated new samples are then passed through the network.
Using the resulting weak predictions and the labels assigned before,
a crossentropy loss for each class is calculated. This loss
indicates how good the weak and strong predictions align.

\end{enumerate}

Afterwards for each class, the post-processing with the smallest
loss value is selected.

 This approach does not need any labels,
 neither strong nor weak. Therefore our method for post-processing selection is applicable
 using data of both, the weakly-supervised and the unsupervised dataset.
 Also the method can be used to adapt the post-processing
 at inference time to new unseen data.

\subsection{Training}

The cross entropy loss between
the predicted probabilities for each class and the weak ground truth labelling
over all labelled clips in a batch
is calculated as follows:

\begin{equation}
E = -\frac{1}{N}
\sum_{i}^N \sum_{c}^M
l_c^{(i)} log(y_c^{(i)}),
\end{equation}

\noindent where the number of classes is denoted by M, the number of weakly
 labelled 10 second audio clips by N,
$y_c^{(i)}$ denotes the predicted probability for class $c$ of sample $i$, and
$l_c^{(i)}$ is the given binary label in the weakly labelled train set.

In each step a batch containing an equal distribution of
samples from the labelled and unlabelled in-domain set is processed.
The total loss consists of
the cross entropy loss of the labelled samples,
regularized with VAT depending on both the labelled and
unlabelled samples weighted by a factor $\lambda$:

\begin{equation}
\begin{multlined}
L =
-\frac{1}{N}
\sum_{i}^N \sum_{c}^M
l_c^{(i)} log(y_c^{(i)})  \\
+ \lambda \sum_i^{N'} \mathrm { KL } [ p (  \mathbf{y}  | \mathbf{x}^{(i)} ; \theta) | | p (  \mathbf{y}  |  \mathbf{x}^{(i)}  + \mathbf{r} ; \theta ) ],
\end{multlined}
\end{equation}

\noindent where $N'$ denotes the sum of labelled and unlabelled in-domain clips in a batch,
$\mathbf{x}^{(i)}$ is the log-scaled mel-spectrograms of a labelled or unlabelled in-domain clip with index $i$.

The loss was optimized using Adam \cite{adam} with a learning rate of $0.001$ and a
batch size of 30.
The network was implemented using tensorflow \cite{abadi2016tensorflow}.

\begin{table*}[h!]
\footnotesize
\centering
\begin{tabular}{l|cc|cc|cc|cc|cc|cc|cc}
\cline{4-15}\multicolumn{1}{c}{} & \multicolumn{2}{c|}{}  &  \multicolumn{6}{c|}{no VAT} & \multicolumn{6}{c}{VAT}  \\ \cline{2-15}
& \multicolumn{2}{c|}{challenge baseline} & \multicolumn{2}{c|}{no refinement} & \multicolumn{2}{c|}{$\text{SALR}_{\text{train}}$} & \multicolumn{2}{c|}{$\text{SALR}_{\text{dev.}}$} &\multicolumn{2}{c|}{no refinement} & \multicolumn{2}{c|}{$\text{SALR}_{\text{train}}$} & \multicolumn{2}{c}{$\text{SALR}_{\text{dev.}}$}\\ \hline
Class         & \textbf{F1} & \textbf{ER} & \textbf{F1} & \textbf{ER} & \textbf{F1} & \textbf{ER} & \textbf{F1} & \textbf{ER} & \textbf{F1} & \textbf{ER} & \textbf{F1} & \textbf{ER} & \textbf{F1} & \textbf{ER} \\ \hline
Alarm bell                                  & 3.2\%     & -         & 27.0\%            & 1.45  & 22.4\%    & 1.18         & 18.8\%            & 1.23              & \textbf{27.9\%}       & 1.38   & 21.0\%   & 1.14       & 18.2\% & \textbf{1.12}   \\
Blender                                     & 15.4\%    & -         & 18.5\%            & 2.65  & 10.7\%    & 1.25         & 26.9\%            & 1.23              & 29.9\%       & 1.52   & 23.2\%   & 1.33       & \textbf{38.1\%} & \textbf{0.97}   \\
Cat                                         & 0.0\%     & -         & 9.5\%             & 3.27  & 5.0\%     & 1.40         & \textbf{33.5\%}             & 1.35              & 4.9\%       & 2.87   & 19.2\%   & 1.54       & 25.2\% & \textbf{1.30}   \\
Dishes                                      & 0.0\%     & -         & 5.6\%             & 1.65  & 0.0\%     & 1.16         & 0.0\%             & 1.16              & 29.3\%        & 1.93   & \textbf{32.5\%}   & \textbf{1.16}       & \textbf{32.5\%} & \textbf{1.16}   \\
Dog                                         & 0.0\%     & -         & \textbf{20.5\%}   & 2.16  & 18.5\%    & 1.40         & 18.6\%            & 1.39              & 7.4\%       & 2.00   & 2.3\%    & \textbf{1.36}       & 15.8\% & \textbf{1.36}   \\
Elec. Shaver                                & 32.4\%    & -         & 18.4\%            & 2.86  & \textbf{50.0\%}    & \textbf{0.86}         & \textbf{50.0\%}            & \textbf{0.86}              & 14.1\%       & 2.61   & 40.0\%   & 0.96       & 40.0\% & 0.96   \\
Frying                                      & 31.0\%    & -         & 20.4\%            & 4.54  & \textbf{43.5\%}    & 1.62         & 42.9\%            & 1.67              & 18.0\%       & 3.79   & 40.0\%   & 1.50       & 40.7\% & \textbf{1.46}   \\
Running water                               & 11.4\%    & -         & 17.5\%            & 1.86  & \textbf{37.7\%}    & 1.00         & 38.0\%            & \textbf{0.99}              & 22.6\%       & 1.89   & 31.1\%   & 1.22       & 32.4\% & 1.21  \\
Speech                                      & 0.0\%     & -         & 36.5\%            & 1.38  & \textbf{44.6\%}    & \textbf{0.95}         & 36.2\%            & 1.15              & 37.5\%       & 1.25   & 41.3\%   & 0.97       & 40.2\% & 0.98   \\
Vac. cleaner                                & 46.5\%    & -         & 20.0\%            & 3.11  & 48.8\%    & 1.17         & 46.5\%            & 1.28              & 21.8\%       & 2.58   & 40.5\%   & 1.31       & \textbf{63.0\%} & \textbf{0.75}   \\ \hline \hline
                                            & 14.06\%   & 1.54      & 19.4\%            & 2.49  & 28.12\%   & 1.19          & 31.2\%            & 1.23              & 21.3\%       & 2.18   & 29.1\%   & 1.25        & \textbf{34.6\%}&  \textbf{1.12} \\
\end{tabular}
\caption{\textbf{Class-wise results} on the \textbf{development set}, total scores
are macro averaged. }
\end{table*}

\section{Experiments and results}
\label{sec:pagelimit}

\subsection{Dataset}
\label{ssec:subhead}

The method is evaluated using a subset of the Google Audioset \cite{gemmeke2017audio},
which was provided with task 4 of the DCASE 2018 challenge\cite{dcase2018web}.

The majority of the provided audioclips are 10 seconds long, a few audioclips
are slightly shorter, for further processing we zero-pad those to a length of 10 seconds.
Each audioclip contains one or multiple sound events of 10 different classes,
whereby different events may overlap.
The dataset consists of a training, testing and evaluation subset.

The training subset consists of 1,578 weakly labelled clips, an unlabelled in-domain set
of 14,412 clips and an unlabelled out-of-domain set of 39,999 clips extracted from classes that are not considered in task 4.

The test set contains 288 clips, whereby the distribution in terms of clips per class is
 similar to the weakly labelled training set. For
 the test set strong labels from human annotators are given, therefore timestamps
 for the onset and offset of each event in the clip are included.
For training only weak labels are used.
The weak labels indicate if a given event occurs somewhere in a 10s clip,
however no information about the onset and offset of the events, nor how
often the event occurs is given.
This setting can also be considered as a multiple instance learning
(MIL) problem \cite{McFee2018AdaptivePO}.

Log-scaled mel-spectrograms of each clip are passed as input to the network, for calculation the librosa library
\cite{mcfee2015librosa} is used.
Before the spectrograms are calculated, each clip is converted to a mono signal with
a sampling rate of 16,000 Hz.
For calculation of the log-scaled mel-spectrograms a hamming window of length 1024 with an overlap of 360 is used, this gives
240 frames with 64 mel frequency channels for each clip.

\vspace{-.3cm}
\subsection{Baseline system}
The organizers of the DCASE challenge provided a baseline system for task 4 \cite{2018arXiv180710501S}.
The system consists of two models based on the same structure:
three convolution layers with 64 filters of size $3\times3$, each one
followed by a max pooling layer of size $4\times1$ and a dropout layer with $p=30\%$.
After the convolutional layers, one bidirectional recurrent layer with 64 GRU units and 30\% dropout on the input
is placed. For output, the first model uses a dense layer with 10 sigmoid units
and global average pooling across frames to make clip-level predictions,
and the second model uses a time distributed dense layer with 10 sigmoid units to
predict events at frame-level.
Training of the system is performed in two steps:

\begin{enumerate}
\item The first model is trained with the weakly labelled training set,
then the trained model is used to generate
weak labels for the unlabelled in-domain set.

\item The second model is trained on the unlabelled in-domain set,
using the weak labels generated beforehand.
In this second training pass the weakly labelled set is used for validation.
 \end{enumerate}

As input, each 10 second audio file is divided into 500 frames of 64 log mel-band magnitudes.

\vspace{-.3cm}
\subsection{Evaluation}

For evaluation the macro averaged event-based F-score \cite{Mesaros2016_MDPI} is used.
The event-based metrics are calculated using the open source toolbox sed\_eval
 \cite{sedeval}.
As given by the challenge,
for calculation of event-based metrics a 200ms collar on onsets and a 200ms / 20\% of the events length collar on offsets
was set.
For calculation of the total performance over all individual classes, macro averaging is used.
This has the effect that each class has equal influence on the final metrics, even if the distribution
of classes in the tested set is unbalanced.

\vspace{-.3cm}
\subsection{Results}
Table 1 shows the event based F1 scores and error rates of our system on the development set.
We compare the resulting scores of our system without post-processing refinement, and when we performed self-adaptive label refinement using data either of the training or the development set.
Additionally, we also show the influence of VAT.
When no post-processing refinement was done, we calculated the strong labels with a fixed
threshold of $0.5$ for all classes and apply no median filter.
It can be seen that both SALR and VAT increase the performance of the system.
Whereby when SALR is used, the best performance is achieved when the adaption was done on the development set.

\vspace{-.3cm}
\subsection{Submitted systems}
Three systems have been submitted to the DCASE 2018 challenge,
whereby self-adaptive label refinement was used to adapt the post-processing as follows:
System one has been adapted to the evaluation set.
System two did not use any adaption, but used the same
post-processing with a fixed threshold of $0.5$ and a median
filter width of 1. System three has been adapted to the training set.

\vspace{-.3cm}
\section{Conclusion}
In this paper, we proposed a method for sound event detection using only weakly labelled and unsupervised data.
Our approach is based on GCRNNs,
whereby we introduce self-adaptive label refinement. This
method adapts the postprocessing using unlabelled data, and increases SED performance.
The final F-score of our system is 34.6\%, which is significantly higher than the score of the baseline system which is 14.06\%.

\bibliographystyle{IEEEtran}
\bibliography{refs}
%
%
%
%
%
%
%
%
%

\end{sloppy}
\end{document}